# Valley Emission and Upconversion in Isotopically Engineered Monolayer WS$_2$ under Resonant Excitation

Rahul Kesarwani[1,2], Vaibhav Varade[1], Artur Slobodeniuk[1], Martin Kalbac[2]*, and Jana Vejpravova[1]*

[1] Department of Condensed Matter Physics, Faculty of Mathematics and Physics, Charles University, Ke Karlovu 5, 121 16 Prague 2, Czech Republic

[2] Department of Low-Dimensional Systems, J. Heyrovsky Institute of Physical Chemistry, Dolejskova 3, 182 23 Prague 8, Czech Republic

*Email: martin.kalbac@jh-inst.cas.cz

*Email: jana.vejpravova@matfyz.cuni.cz

Keywords: 2D materials, isotope engineering, resonant excitation, valley polarization, photoluminescence, upconversion, interband and intraband scattering.

**Abstract**

In the quest to optimize the optoelectronic and valleytronic properties of 2D materials, various strategies such as strain engineering, doping, and heterostructuring have been explored. In this direction, isotope engineering also offers a potential avenue to alter electron-phonon interaction and impact quasiparticle scattering processes. In this study, we investigate the dependence of sulfur isotopes on upconversion and valley scattering phenomena by collecting the resonance photoluminescence (PL) under an applied magnetic field from 0 to 14 T at 4 K for the chemical vapor deposition-grown monolayer (1L) of W$^N$S$_2$, W$^{32}$S$_2$, and W$^{34}$S$_2$. The upconversion of the mixed-state sulfur 1L (W$^N$S$_2$) exhibits one $M$-phonon absorption, with an obtained optical gain of nearly 30 meV, while the pure sulfur isotope labelled 1Ls (W$^{32}$S$_2$ and W$^{34}$S$_2$) require two phonons ($M$ and $\Gamma$), yielding a gain of around 80 meV. It is also found that the exciton degree of polarization (DOP) of W$^N$S$_2$ changes significantly by ~ -30% as the field increases from 0 to 14 T, while for W$^{32}$S$_2$ and W$^{34}$S$_2$, the exciton DOP increases by up to ~8%. Similarly, distinct changes in the DOP are observed for trions and localized excitons among all the samples, attributed to the different valley scattering phenomena. The 1L W$^N$S$_2$ demonstrates a combination of intraband and interband scattering, whereas in the case of W$^{32}$S$_2$ intraband scattering is preferred; W$^{34}$S$_2$ predominantly exhibits interband scattering. Finally, a phenomenological model is proposed to describe the upconversion and valley scattering processes.

# 1. Introduction

Two-dimensional (2D) transition metal dichalcogenide (TMD) materials have emerged as promising platforms for various concepts for optics, sensing, and nanoelectronics [1–3]. Recently, upconversion phenomena have been reported as a promising approach to boost the opto-electronic properties of TMDs, wherein lower-energy photons are efficiently converted into higher-energy ones [4,5]. By converting lower-energy photons into higher-energy ones, upconversion transitions enhance the overall efficiency of light emission, which is advantageous for applications such as solar cells, light-emitting diodes, optoelectronic devices, and optical refrigeration in semiconductors [6–9]. Upconversion mechanisms, including excited state absorption, sequential energy transfer, cooperative transitions, and photon avalanche, are governed by distinct electronic interactions. Among these, energy transfer and excited state absorption dominate due to their higher efficiencies, especially in sensitizer–activator systems [10]. Understanding stepwise electronic excitation, phonon-assisted energy transfer, and non-radiative relaxation is crucial for optimizing upconversion materials. Leveraging the unique properties of 2D-TMD materials, including their atomically thin nature (enabling enhanced quantum confinement and reduced dielectric screening) and tunable electronic structure (facilitating direct-to-indirect bandgap transitions and strong excitonic effects), various mechanisms have been proposed, including electron-electron, electron-phonon, exciton-phonon, and phonon-phonon interaction processes, potentially influencing parameters such as the degree of polarization (DOP), intra/interband scattering, binding exciton energy, trion-to-exciton ratios, and localized exciton states [5,11–15]. Understanding and optimizing these processes is crucial for enhancing the efficiency and applicability of upconversion in 2D-TMD materials.

Furthermore, upconversion transitions enable the detection of weak signals by amplifying them through the accumulation of photons during the absorption process. This heightened sensitivity is invaluable for various sensing and imaging applications, including biological sensing and environmental temperature monitoring [16,17]. In monolayer (1L) TMDs, this phenomenon can lead to unique optical properties and novel applications, such as nonlinear optics and quantum information processing [18–20]. More significantly, photoluminescence (PL) in 1L TMDs can potentially be regulated by isotope engineering (IE) (substituting transition metal (TM) and sulfur (S) isotopes), electrical doping, charge transfer through heterostructures, structural light, etc. [21–

26]. This alteration can further impact the properties of electron states in the crystal, leading to the modification of the electronic band structures, resulting in the modification of the exciton masses and binding energies [27,28]. Furthermore, Wu et al. [29] and Vuong et al. [30], respectively, discussed the theoretical influence of tungsten (W) isotopes and experimentally exhibited the effect of boron (B') isotopes on spin-dependent properties and intra/interband transitions/scattering in both $K^+$ and $K^-$ valleys. Such isotopic effects offer a novel opportunity for tuning 2D materials' physical properties while maintaining an identical chemical composition. Likewise, the isotopic effect of S on optical properties in TMDs has been minimally explored. Early discussions by Varade et al. [31] emphasized the S isotopic effect on the optical properties (Raman and PL) of 1L $MoS_2$ under non-resonance excitation, showing the fundamental role of S isotopes in the phonon-driven properties of $MoS_2$.

Understanding the impact of S isotopes on upconversion phenomena under resonance conditions could complete the picture of valley scattering processes. Earlier studies have focused on oxide isotope precursors and magnetic field effects, while S isotopes may offer additional insights into these dynamics. Although reports on the effect of magnetic field on resonance PL in 1L TMDs are limited, it is well-established that magnetic fields enhance quasiparticle scattering between the $K^+$ and $K^-$ valleys [32–34]. However, similar scattering processes have not been widely discussed for other material systems beyond 2D TMDs. Therefore, further comprehensive investigations are required to analyze the isotopic effects of S on the Raman and PL properties, under resonance excitation by varying magnetic fields and temperatures, of 1L TMDs.

Over the past decade, researchers have expanded the scope of IE to encompass graphene and various semiconductors [35,36]. While materials like $WS_2$ and $MoS_2$ primarily involve oxide isotope variations and hBN focuses on B' isotopes, detailed reports on IE in 2D materials remain limited [21,30,31,37–39]. The works predominantly focus on varying TM isotopes (W, Mo, and B') and their impact on transport properties, limiting the investigations of the optical properties, which are essential with respect to the most promising applications of TMDs. Additionally, it has been reported that the alteration of isotopes can modify both electron-phonon interactions and the exciton dynamics [40]. Despite generic claims of the minimal effects of the chalcogenide isotopes (S) on physical properties (both optical and transport), experimental and theoretical evidence remains inconclusive [29].

The present study investigates the isotopic effects of S in 1L WS$_2$, with a focus on PL under resonant excitation. The chemical vapor deposition (CVD) technique is employed to fabricate high-quality WS$_2$ monolayers incorporating natural S, $^N$S (W$^N$S$_2$), and isotopically pure $^{32}$S (W$^{32}$S$_2$) and $^{34}$S (W$^{34}$S$_2$), recognized as the most stable and prevalent isotopes within the S isotopic spectrum. The polarization-resolved optical experiments are carried out down to cryogenic temperatures under magnetic fields ranging from 0 to 14 T. Our results stress a significant influence of IE on PL upconversion and make it possible to uncover the interband and intraband scattering mechanisms under the applied magnetic field. The study provides important insights into the intricate interplay of isotopic effects in valley-driven physics and upconversion processes in 1L TMDs.

## 2. Experimental Details

WS$_2$ 1Ls were grown on a silicon (Si) substrate covered with a 300-nm-thick layer of silicon oxide (SiO$_2$). The substrate was cleaned through subsequent sonication in deionized water, acetone, and isopropanol (Sigma-Aldrich). The fabrication of WS$_2$ took place in a quartz tube using a horizontal furnace with two heating zones, under atmospheric pressure, employing argon as the carrier gas. Initially, the tube was connected to an argon gas line at one end and to a bubbler filled with a 100 mM aqueous solution of KOH. The tube was flushed with argon at a continuous flow of 200 cm$^3$ min$^{-1}$ at a temperature of ~105°C for 15 minutes. The precursor, WO$_3$ (155 mg) powder (Thermo Scientific—LOT: R04G047), was placed in the crucible at the first heating zone (high temperature), and the substrate was placed face-down on top of the crucible. Simultaneously, 100 mg of $^N$S (Thermo Scientific—LOT: T29G013) was placed in a tube, 20 cm away, at the second heating zone (low temperature). Maintaining a constant argon flow rate of 150 cm$^3$ min$^{-1}$, the temperature was subsequently increased in the tube at a constant rate of 40°C min$^{-1}$. When the temperature reached ~770°C in the first heating zone, the heating in the second zone was initiated. The temperatures were held constant at ~990°C in the first zone and ~220°C in the second zone for 10 minutes, after which the furnace was opened, and the system was allowed to cool ambiently. A similar process was adopted for the other two isotopes of sulfur to obtain 1Ls of W$^{32}$S$_2$ and W$^{34}$S$_2$.

Atomic force microscopy (AFM) images and thickness profiles were obtained using Bruker's AFM Dimension ICON system in the Peak Force Quantitative Nanomechanical Mapping mode with a Bruker silicon tip. The AFM data were processed and analyzed by open-access Gwyddion software.

Micro-Raman and PL spectra were measured using an in-house spectroscopic setup that collects the signals in a backscattering geometry using a low-temperature confocal Raman microscope insert (attoRAMAN, attocube) placed in a Physical Property Measurement System (PPMS, Quantum Design). A low-temperature- and magnetic-field-compatible 100× objective (numerical aperture 0.82 and lateral resolution of 500 nm) lens was used to focus the resonance (633 nm or 1.96 eV) and non-resonance (532 nm or 2.33 eV) excitation with linear and circularly polarized light at a power of 100 µW. The WITec spectrometer was used to scan the map of 1L flakes of size 20–30 µm for PL and Raman spectra with a spatial resolution of 0.5 µm using a grating of 600 and 1800 lines/mm, respectively. The incident laser beam was circularly polarized using a set of standard 633 nm half- and quarter-wave plates. Similarly, a series of broadband quarter- and half-wave plates were used to obtain polarization-resolved emitted signals. More details regarding the experimental setup are discussed in our previous paper [41].

## 3. Results and Discussion

### Raman and PL Characterizations

We prepared samples of $W^NS_2$, $W^{32}S_2$, and $W^{34}S_2$ using CVD method from isotopically labeled precursors. AFM and Raman analysis confirm that the prepared $WS_2$ samples consist solely of monolayers. The AFM images and its thickness measurements are shown in Figure 1.

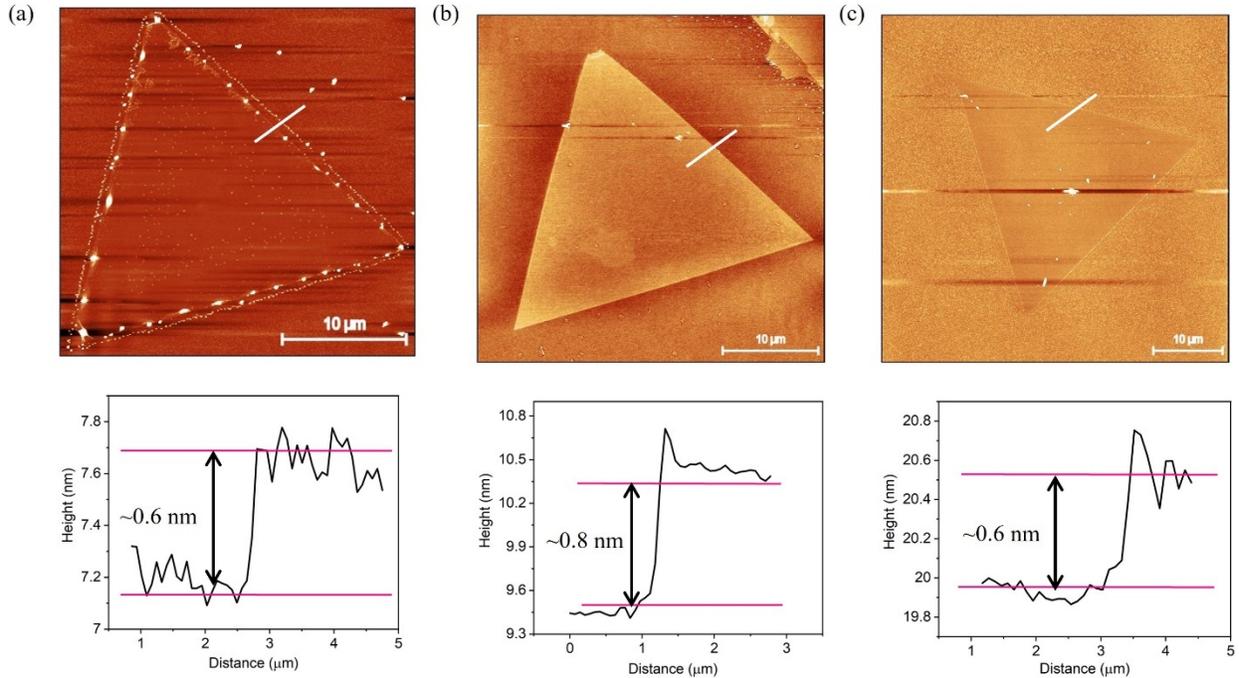

Figure 1. Surface morphology of isotope modified monolayers (1L) WS$_2$. AFM analysis for surface profile and thickness measurement of 1-Ls. (a) W$^N$S$_2$, (b) W$^{32}$S$_2$, and (c) W$^{34}$S$_2$.

The Raman spectra of isotopically labeled WS$_2$ monolayers at 300 K (under non-resonant 532 nm, 2.33 eV excitation) are shown in Figure 2(a). We observe that the peak position of the A$_{1g}$ phonon mode remains nearly identical for W$^N$S$_2$ (mixer of S isotopes: $^{32}$S (95%), $^{34}$S (4.25%) and $^{33}$S (0.75%)) and W$^{32}$S$_2$, while it is redshifted in W$^{34}$S$_2$ (inset of Figure 2(a)). This redshift primarily results from the S isotope mass effect, where the heavier $^{34}$S isotope leads to a decrease in phonon energy. The phonon frequency follows an inverse square root dependence on mass ($\omega \propto 1/\sqrt{m}$), leading to a systematic shift in the Raman-active modes of isotopically engineered materials [42]. Although strain and phonon dispersion variations can influence Raman shifts, the fewer shift between W$^N$S$_2$ and W$^{32}$S$_2$ suggests that isotope-induced mass changes are the dominant factor. The observed shift is consistent with previous reports on isotopically modified TMDs [31]. Figure 2(b) shows the deconvoluted Raman spectrum of W$^{32}$S$_2$ at 4 K (under excitation 633 nm, 1.96 eV). In this case, additional phonon modes are activated, and new bands are observed at 385 cm$^{-1}$, 355 cm$^{-1}$, 213 cm$^{-1}$, 171 cm$^{-1}$, and 145 cm$^{-1}$ in addition to the A$_{1g}$ and E$_{2g}$ modes. The phonons of resonant modes can be responsible for upconversion and intra/interband phenomena (further details discussed below) [43]. Full deconvoluted Raman spectra measured at the resonance excitation of W$^N$S$_2$ and W$^{34}$S$_2$ are presented in Figure S1 (SI). The excitation energy is not expected to affect the Raman peak positions. We indeed observed a similar difference in the shift of the $A_{1g}$ phonon mode for the $W^NS_2$, $W^{32}S_2$, and $W^{34}S_2$ 1Ls (see Table S1 in SI), as in the case of non-resonance conditions. (The complete list of the resonance phonon peaks, along with their respective assignments for all the isotopes of modified $WS_2$, is shown in Table S1 (SI).)

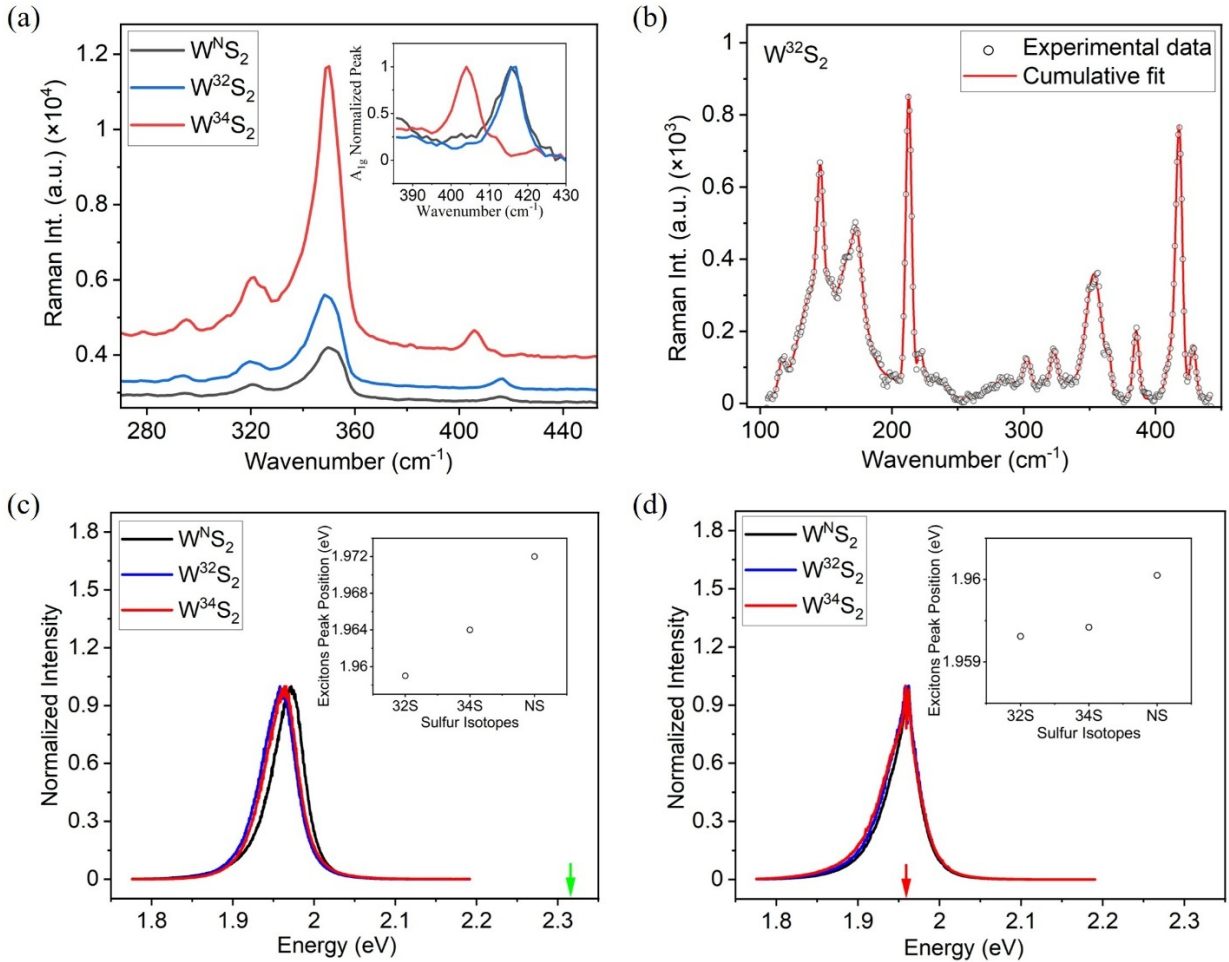

Figure 2. Raman/PL of the $W^NS_2$, $W^{32}S_2$, and $W^{34}S_2$ 1Ls: (a) Raman spectra at 300 K under non-resonance excitation. (b) Raman spectra at 4 K under resonance excitation. PL spectra at 300 K under (c) non-resonance and (d) resonance excitation. The arrow represents the excitation energy.

The PL spectra measured for $W^NS_2$, $W^{32}S_2$, and $W^{34}S_2$ at 300 K are shown in Figure 2(c) and (d) under non-resonance and resonance excitation, respectively. Upon initial inspection, the energy bandgap (or excitonic peak) of isotope-modified $WS_2$ for downconversion (non-resonance excitation, inset of Figure 2(c) shows a blueshift of ~ 4 meV with varying isotopes from $^{32}S$ to $^{34}S$, as observed in conventional semiconductors due to the involvement of phonons in the indirect band-to-band transition [30,44]. Further, the exciton energy is found to blueshift by ~7 meV in $W^NS_2$ with respect to $W^{32}S_2$, which can be attributed to isotope mixture-induced disorder. In contrast, in the case of upconversion (resonance excitation, inset of Figure 2(d)), the exciton energy shift due to different isotopes is negligible due to direct band-to-band transition, which does not involve phonons.

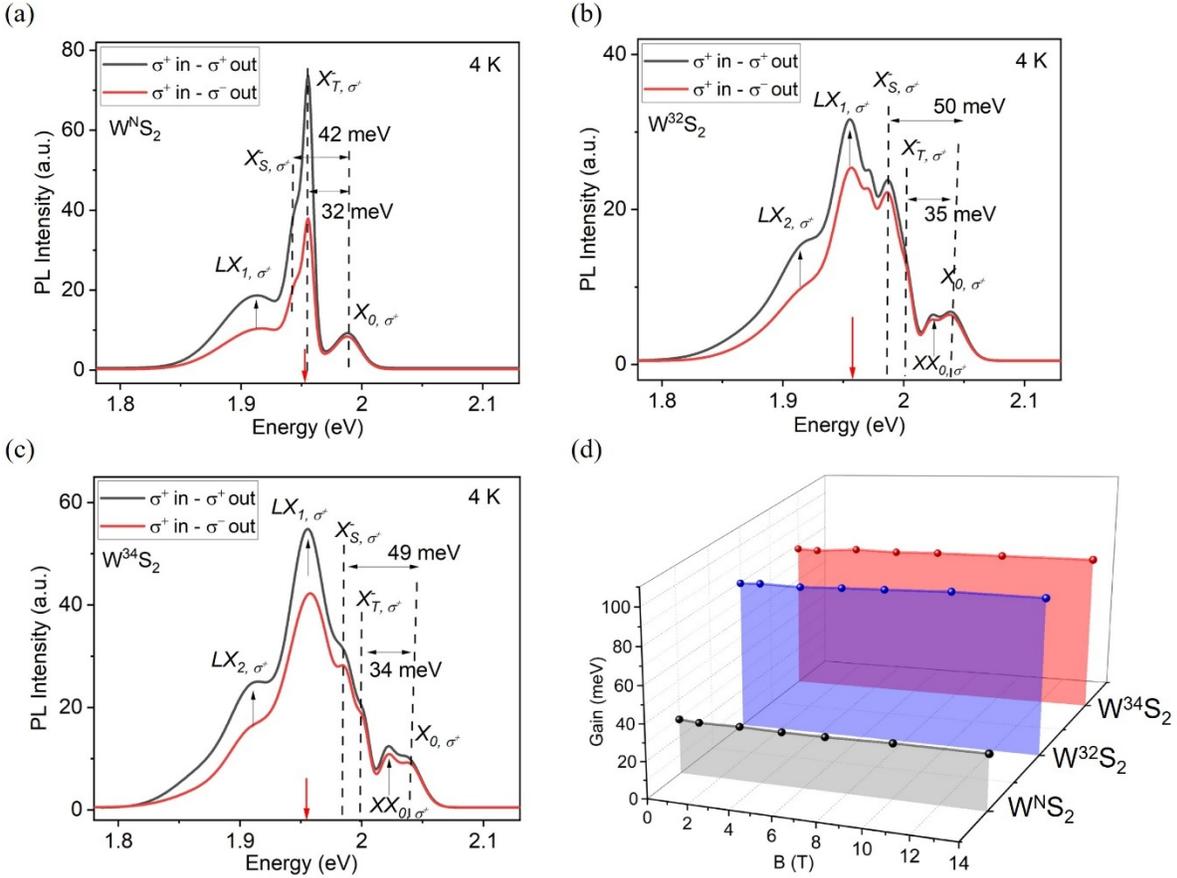

Figure 3. PL spectra of sulfur isotopes of modified $WS_2$ 1Ls at 4 K and 0 T under resonance excitation: (a) $W^NS_2$, (b) $W^{32}S_2$, and (c) $W^{34}S_2$. The assigned peaks of exciton ($X_{0,\sigma^+}$), trions ($X^-_{i(=T,S),\sigma^+}$), and localized excitons ($LX_{i(=1,2),\sigma^+}$) are identified after the deconvoluted PL spectra of $\sigma^+$ excitation ("$\sigma^+$ in") and $\sigma^+$ detection ("$\sigma^+$ out"), along with their binding energy. (d) Comparison of exciton gain among different isotopes is obtained using "$\sigma^+$ out" polarization state depending on applied magnetic fields. The arrow indicates the excitation energy.

To probe the circular DOP and intra/interband scattering within/between $K^+$ and $K^-$ valleys, we collected circularly polarized magneto-PL spectra at 4 K for the set of isotopically modified WS$_2$ 1Ls. The magneto-PL measurements are carried out with the excitation of right circularly polarized light, denoted as "$\sigma^+$ in," and the emission is detected in both the right and left circular polarization states, referred as "$\sigma^+$out" and "$\sigma^-$out," respectively, under varying magnetic fields from 0 to 14 T. Figure 3(a)–(c) shows the initial PL spectra of both "$\sigma^+$out" and "$\sigma^-$out" for the 1Ls of W$^N$S$_2$, W$^{32}$S$_2$, and W$^{34}$S$_2$, respectively, recorded at 4 K at 0 T. The PL of 1L WS$_2$ is characterized by distinct quasiparticles, and their identified peak positions for "$\sigma^+$out" polarized light are shown. These PL bands are assigned as follows: "exciton" refers to "$X_{0,\sigma^+}$," "biexciton" to "$XX_{0,\sigma^+}$," "triplet trion" to "$X^-_{T,\sigma^+}$," "singlet trion" to "$X^-_{S,\sigma^+}$," and "localized excitons" to "$LX_{i\,(=1,2),\sigma^+}$." Notably, the characteristic PL spectra at 4 K of the 1Ls W$^{32}$S$_2$ and W$^{34}$S$_2$ exhibit two $LX$ peaks and one neutral $XX_0$ peak, while the mixed-isotope 1L shows only one $LX$ peak. The complete deconvoluted PL spectra of both the "$\sigma^+$out" and "$\sigma^-$out" polarization states of all the isotopes of modified WS$_2$ measured at 4 K are shown in Figures S2 and S3 (in the SI). The peak positions obtained by deconvoluting the PL spectra are listed in Table S2 in the SI; for example, $X_{0,\sigma^+}$, $X^-_{T,\sigma^+}$, and $LX_{1,\sigma^+}$ for W$^N$S$_2$ at 4 K and 0 T were found to be around 1.989 eV, 1.956 eV, and 1.913 eV, respectively, and they are in agreement with the previous reports on CVD-grown 1L WS$_2$ [5,12,45–47]. Spectral deconvolution of the PL spectra was carried out using Origin software, where each spectrum was fitted with Gaussian peaks to determine peak positions, area under the curve, and intensity ratios for various quasiparticles. The best fit was selected based on R$^2$ values (close to 1), ensuring an accurate and reliable decomposition of the spectra.

Importantly, S isotopes impact the optical properties of 1L WS$_2$, as evidenced by the characteristic PL spectra and estimated trion binding energies. The estimated binding energies ($E_{bX^-_{i(=T,S)}} = X_0 - X^-_{i\,(=T,S)}$) of $X^-_{T,\sigma^+}$ and $X^-_{S,\sigma^+}$ in W$^N$S$_2$ are 32 and 42 meV, respectively, which results in an energy difference ($\delta_{TS}$) between $X^-_{T,\sigma^+}$ and $X^-_{S,\sigma^+}$ of 10 meV, consistent with previous reports [5,46,47], where the values were observed from 6 to 11 meV. However, in W$^{32}$S$_2$, the binding energies for $X^-_{T,\sigma^+}$ and $X^-_{S,\sigma^+}$ are 35 and 50 meV, respectively, resulting in $\delta_{TS}$ ~15 meV. Similarly, for W$^{34}$S$_2$, the binding energies are nearly identical (34 and 49 meV) to W$^{32}$S$_2$, with the same $\delta_{TS}$ (displayed in Figure 3). $\delta_{TS}$ is found to increase slightly (2–3 meV) with varying magnetic fields (0 to 14 T) in the case of all the isotope-modified WS$_2$ 1Ls; see Table S2 in the SI. Some reports indicated that

$\delta_{TS}$ decreases if an hBN spacer is placed between WS$_2$ and the substrate [5,11]. Building upon this finding, the observed variations in both the PL characteristic and binding energies within isotope-modified WS$_2$ samples can be ascribed to a S lattice mismatch with the substrate lattice. In the case of the mixed state (W$^N$S$_2$), the S lattice mismatch is likely to be more pronounced, leading to reduced interaction with the substrate. Conversely, the pure isotope states of samples demonstrate a strong interaction with the local field of the SiO$_2$ substrate, resulting in the induction of a more localized band and thus increased $\delta_{TS}$ [11,48,49]. To further investigate the influence of substrate interactions and strain effects, we analyzed the correlation between the E$_{2g}$ and A$_{1g}$ Raman modes. The observed shifts provide insights into possible strain contributions, supporting the hypothesis of substrate-induced effects in the pure isotope samples. A correlation plot between the E$_{2g}$ and A$_{1g}$ Raman mode shifts is presented in Figure S4 (in SI).

To further verify the positions of the quasiparticle peaks in the case of isotopically pure samples, we analyzed the temperature dependence of PL for the "$\sigma^+$out" detection of 1L W$^{32}$S$_2$. The observed changes of peak shifts with temperature, as illustrated in Figure S5(a), align with previously reported behavior [45,50]. Additionally, the positions of excitons and trions versus temperature closely follow the Varshni model (Figure S5(b) and (c)), while localized excitons fit well with the Manoogian model (refer to Figure S5(d)) [31,51]. Furthermore, we analyzed the integral intensity ratio, $I_{X^-_{\sigma^+}}/I_{X_{0,\sigma^+}}$, of trions to excitons (Figure S5(e)); it shows that trions are more pronounced at low temperatures due to reduced thermal energy and phonon coupling effects [52]. Further, Figure S5(f) shows the intensities of quasiparticles at different temperatures following the usual trend in resonant excitation reported previously [51,53]. The power dependence of the intensity of excitonic peaks at 4 K and 300 K (Figure S5(g)) is found to follow a typical power law with an exponent value ~ 0.77 [50]. Moreover, the neutral $XX_0$ peak is observed only in the 1Ls W$^{32}$S$_2$ and W$^{34}$S$_2$. The peak assignment of $XX_0$ is confirmed by analyzing the ratio of the scaling factor between the assigned peak and the exciton peak intensity variation with excitation power. It was found that the scaling factor for the biexciton is twice that of the exciton, confirming that the assigned peak in our PL spectra corresponds to biexcitonic emissions (see Figure S5(h) in SI). This analysis also highlights the spectral variations between pure and mixed isotopes. Additionally, Figure S5(i) – (j) illustrates the variation of trion binding energy with excitation power, showing an increase at 300 K that stabilizes at higher power levels, while at 4 K, the variation remains minimal (1–2 meV), consistent with previous reports [26,54].

The optical gain G is defined as the difference between the incident light energy (1.962 eV) and exciton energy, $X_0$ (G = 1.962 – $X_0$). A positive value of G indicates upconversion phenomena, while a negative value denotes downconversion [55]. Upconversion phenomena have been observed for 2D materials under resonance excitation in earlier reports [5,12,13]. Figure 3(d) illustrates the optical gain obtained from measurement of the "$\sigma^+$out" PL signal for $W^NS_2$, $W^{32}S_2$, and $W^{34}S_2$, with the magnetic field being varied from 0 to 14 T. We observed that the optical gain for $W^NS_2$ is approximately 30 meV, while for $W^{32}S_2$ and $W^{34}S_2$, it is around 80 meV for a 0 T field and remains nearly constant as the magnetic field is increased to 14 T for all the measured samples. This can be explained by the fact that the Zeeman shift of exciton energies is an order of magnitude smaller (typically 2 meV per 10 T) than the observed shifts of tens of meV [56]. Additionally, the observed variation in optical gain among the samples likely arises from differences in excitonic interactions and phonon-mediated processes influenced by the isotope mass effect. Similar behavior has been observed for the "$\sigma^-$out" detection polarization states of $WS_2$ 1Ls irrespective of their S isotopic composition, as shown in Figure S6 (SI).

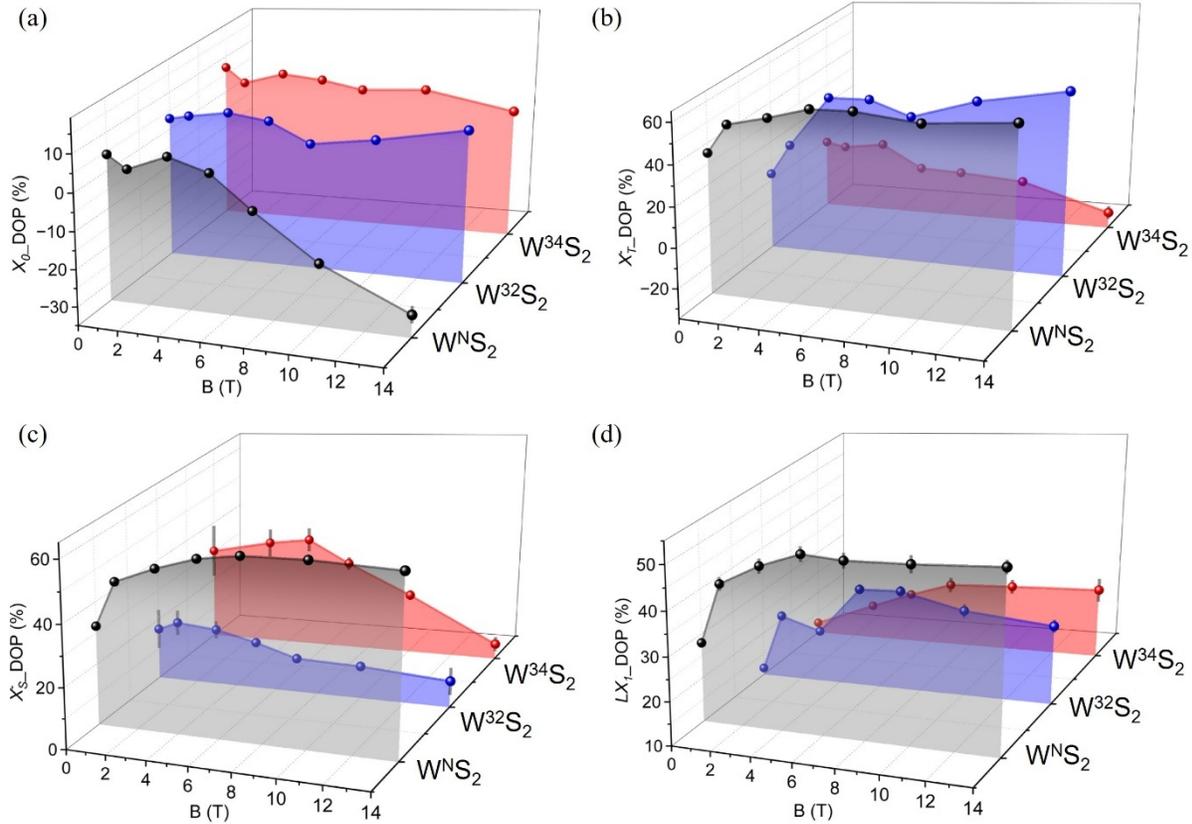

Figure 4. DOP of S isotopes of modified WS$_2$ 1L at 4 K under magnetic field: (a) Exciton ($X_0$), (b) triplet trion ($X_T^-$), (c) singlet trion ($X_S^-$), and (d) localized exciton ($LX_1$).

**DOP and Intensity Ratios**

The circular DOP is defined as $DOP = \frac{I_{\sigma+} - I_{\sigma-}}{I_{\sigma+} + I_{\sigma-}}$, where $I_{\sigma+}$ and $I_{\sigma-}$ represent the intensities of quasiparticles extracted from the deconvoluted PL spectra of "$\sigma^+$out" and "$\sigma^-$out" detection, respectively (note that all the measurements are performed using "$\sigma^+$in" excitation) [23]. The estimated "$X_0$ DOP" percentage as a function of magnetic fields for all studied samples is shown in Figure 4(a). In the case of 1L W$^N$S$_2$, "$X_0$ DOP" changes from 4% to -30%, with the field increasing from 0 to 14 T, while for pure sulfur isotopes of modified W$^{32}$S$_2$ and W$^{34}$S$_2$, the DOP changes from 3% to 0% and 7% to -2%, respectively. Furthermore, Figure 4(b) and (c) displays the calculated "$X_T^-$ DOP" and "$X_S^-$ DOP" as a function of a magnetic field. It is observed that "$X_T^-$ DOP" for 1L W$^N$S$_2$ initially increases from 35% to 53% as the field varies from 0 to 5 T and then saturates at higher fields. However, for pure isotope-modified W$^{32}$S$_2$, "$X_T^-$ DOP" changes from 3% to 51%, and interestingly, for W$^{34}$S$_2$, it changes from 0% to -27% with a field varying from 0 to 14 T. "$X_S^-$ DOP" for W$^N$S$_2$ initially rises from 32% to 55% with a field varying from 0 to 5 T and then saturates at higher fields. However, for pure isotopes, W$^{32}$S$_2$ and W$^{34}$S$_2$, "$X_S^-$ DOP" varies from 17% to 8% and 31% to 4%, respectively, as the field changes from 0 to 14 T. Figure 4(d) shows "$LX_1$ DOP" for all samples as a function of the applied magnetic field. In this evaluation, we observed that for 1L W$^N$S$_2$, "$LX_1$ DOP" initially increases from 27% to 48% as the field increases from 0 to 5 T, and thereafter it saturates at higher fields. For the pure S isotope-modified 1Ls W$^{32}$S$_2$ and W$^{34}$S$_2$, the DOP is increased from 12% to 28% and from 11% to 24%, respectively, as the magnetic field increases from 0 to 14 T. Notably, "$LX_1$ DOP" demonstrates a rapid increase initially with the application of the field (0 $\rightarrow$ 1 T), followed by a slower rate of increase. "$LX_2$ DOP" for W$^{32}$S$_2$ and W$^{34}$S$_2$ changes from 19% to 58% and 32% to 43%, respectively, with the field ranging from 0 to 14 T (see SI Figure S7). Conversely, "$XX_0$ DOP" changes very minimally with the applied fields (see SI Figure S7). The observed changes of the DOP with the magnetic field can be attributed to interband/interband scattering mechanisms, whose influence has been quantified through the analysis of intensity ratios. Further elucidation on this matter is provided in the next section.

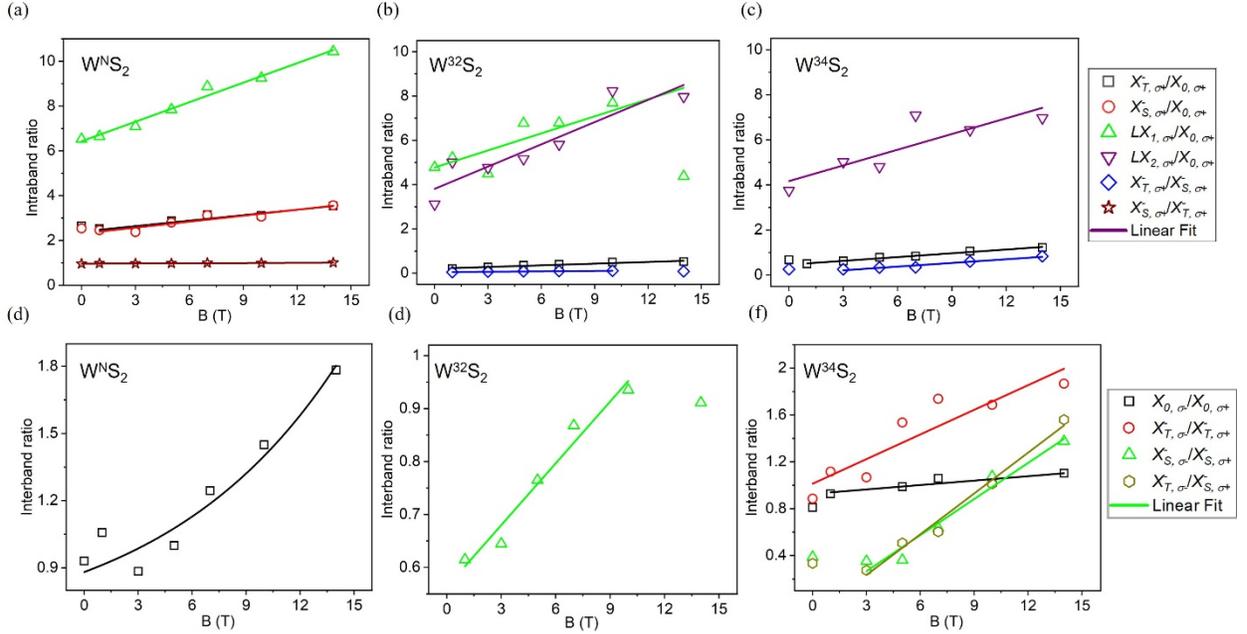

figure 5. Integral intensity ratios of the highly probable band scattering of quasiparticles under magnetic field ranging from 0 to 14 T at 4 K: (a–c) Intraband ratios and (d–f) interband ratios for $W^NS_2$, $W^{32}S_2$, and $W^{34}S_2$, respectively. Favorable conditions for quasiparticle scattering are indicated by positive slopes of the ratios shown in panels (a) to (f).

Figure 5(a) – (c) shows the intraband intensity ratios estimated between different quasiparticles observed in 1L $WS_2$ spectra across an applied field ranging from 0 to 14 T at 4 K for $W^NS_2$, $W^{32}S_2$, and $W^{34}S_2$, respectively. The integral intensity values are derived from resonance PL spectra (as shown in Figure 3 and Figure S3(a) in the SI), capturing PL emission and absorption solely from the transitions within the $K^+$ valley. Earlier studies have indicated that the applied magnetic field can induce alterations in the nature and density of quasiparticles, such as intraband ($K^+ \leftrightarrow K^+$) and interband ($K^+ \leftrightarrow K^-$) scattering, for example, between $X_0 \leftrightarrow X_T^-$ or $X_0 \leftrightarrow LX_1$ [32,57]. The scattering efficiency/probability can be determined by relative intensity ratios (like $X_T^-/X_0$, $X_S^-/X_0$, $X_T^-/X_S^-$, $LX_1/X_0$, etc.), which correlate to the relative amount of emission of quasiparticles with respect to $X_0$, $X_T^-$, $X_S^-$, and $LX_i$ ($i = 1, 2$) within the $K^+$ valley. The intraband ratios for 1L $W^NS_2$, $W^{32}S_2$, and $W^{34}S_2$, plotted against a field, have been fitted by a linear function (*here only positive slope ratios are shown*). The positive sign of the slopes (*m*) determines the explicit scattering mechanism (like

$X_0 \rightarrow X_T^-$, $X_0 \rightarrow LX_1$, etc.), while a slope of zero indicates the absence of changeable phenomena [26].

Figure 5(d) −(f) illustrates the interband quasiparticle peak intensity ratios for 1L $W^NS_2$, $W^{32}S_2$, and $W^{34}S_2$, respectively, under applied magnetic fields at 4 K. These intensity ratio values are determined from resonance PL spectra, as depicted in Figure 3 and Figure S3(b) in the SI, acquiring the absorption from the $K^-$ valley. Notably, in Figure 5(d), it is demonstrated that the interband intensity ratio of 1L $W^NS_2$ (or the mixed state) exhibits an exponential variation with the applied field, whereas 1L $W^{32}S_2$ and $W^{34}S_2$ show a linear change. This proves that the hyperfine interaction induced by the transition metal is not solely responsible for variation in valley scattering; phonon dispersion induced by sulfur isotope precursors can also provoke substantial platforms for intraband and interband scattering, contrary to a previous report [29]. Consequently, changes in S isotopes lead to variations in phonon dispersion, which refers to how the energy and momentum of phonon modes relate to each other [39,58–60]. This alteration affects the phonon spectrum (see Figure 2(b) and Figure S1), influencing the intra- and inter-valley scattering behavior. The fitted slopes for all possible intraband and interband ratios are detailed in Table S3 (SI).

**Intraband and Zeeman Splitting**

Further, we have estimated the intraband (band splitting in the $K^+$ valley) and Zeeman splitting (relative splitting between the $K^+$ and $K^-$ valleys) of 1L $W^NS_2$, $W^{32}S_2$, and $W^{34}S_2$ (see Figure S8). The intraband shifting of trions with respect to excitons indicates a blueshift for the pure S isotope 1Ls $W^{32}S_2$ and $W^{34}S_2$, while the isotopically mixed sample ($W^NS_2$) exhibits a redshift as the magnetic field increases from 1 to 14 T. In the case of 1L $W^NS_2$, this discrepancy can be observed due to increased disorder in $W^NS_2$ and a higher density of defects in the layer lattice compared to pure isotope samples (1L $W^{32}S_2$ and $W^{34}S_2$.) The presence of disorder induced by isotopic mixing in the mixed-isotope $W^NS_2$ system can significantly impact various factors, including the electronic structure, hyperfine interactions [29], and phonon dispersion [39,59]. These disruptions exacerbate differences in exciton and trion state behavior under a magnetic field, ultimately leading to the observed contrasting spectral shifts [46,61]. Furthermore, the extent of Zeeman splitting in the applied magnetic field corresponds to the value of the g-factor. The valley quasiparticle ($X_0$, $X^-_{i=T,S}$, etc.) Zeeman splitting, which is the difference between the quasiparticle transition energies in the $K^+$ and $K^-$ valleys, is defined as $\Delta E \equiv E(K^+) - (K^-) = -g\mu_B B$, where $g$ denotes the $g$- factor of the corresponding excitonic complex and $\mu_B$ is the Bohr magneton. In the case of the 1L $W^NS_2$ exciton, it is reported that the splitting is dominated by the atomic orbital moment of the top valence band at the $K^+$ and $K^-$ valleys, whose g-factor is ≈ 4 [33,62,63]. The g-factor of a negatively charged trion varies from 3 to 5, which was explained by the combined contributions of the spin alignment, the Berry curvature, and the atomic orbital moment [62,63]. However, in the present case, the obtained absolute values of the g-factor for the exciton are 4, 2.3, and 1.7, while for trions, the g-factor values are found to be 3.06, 10, and 2.91 for the 1Ls of $W^NS_2$, $W^{32}S_2$, and $W^{34}S_2$, respectively. (The full list of quasiparticle g-factors is shown in the SI in Table S5.) The g-factor values of excitons and trions for $W^NS_2$ are matched with the reported values [33,63,64]. However, there are no reports for the g-factor values of $W^{32}S_2$ and $W^{34}S_2$. The variation in the g-factor among these samples may stem from reduced atomic orbital magnetic moments as the mass of pure S isotopes increases, alongside alterations in the Berry curvature distribution. The Berry curvature notably impacts phonon dispersion differently across isotopes, influencing the vibrational mode topology and thereby leading to anomalous phonon transport phenomena in the materials [65]. Moreover, for localized excitons exhibiting significant splitting or large g-factors, the underlying mechanisms remain unclear [34]. All fitted slopes for the relative difference between intraband and estimated quasiparticle g-factors are listed in Tables S4 and S5, respectively, in the SI.

**Valley Scattering and Upconversion: A Phenomenological Approach**

We propose a phenomenological model, as shown in Figure 6, to explain the resonance PL observed in Figure 3 and Figure S3 and the variation in its DOP with an applied magnetic field at 4 K (Figure 4), for the 1Ls of $W^NS_2$, $W^{32}S_2$, and $W^{34}S_2$. The model is based on the results obtained from optical gain (Figure 3(d)) and intensity ratio analysis (Figure 5). The resonance magneto-PL can be described by the two-step process: the first step involves upconversion phenomena due to resonance excitation without a magnetic field, and the second step involves intraband/interband scattering in the $K^+$ and $K^-$ valleys upon the application of a magnetic field [12,32]. The valleys in 1L TMD are highly sensitive to magnetic fields and light polarization properties [32,57,66,67]. The below-described model is developed for applied fields ($B > 0$) and resonance excitation under light conditions of "$\sigma^+$ in."

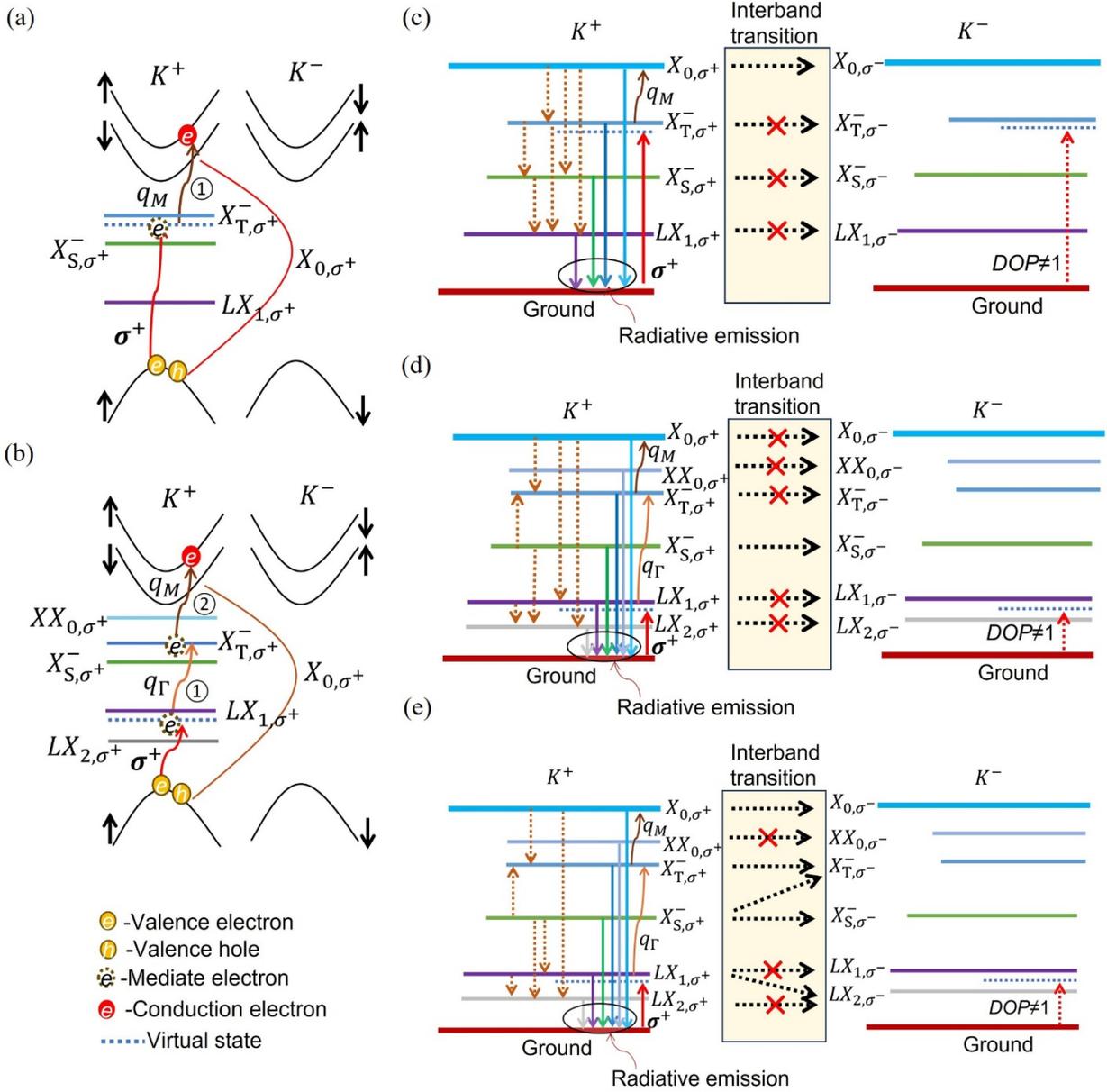

Figure 6. Phenomenological model for resonance magneto-PL of S isotope-modified 1L $WS_2$ at 4 K: (a) and (b) Distinct energy band diagrams for 1Ls of mixed ($W^NS_2$) and pure sulfurs ($W^{32}S_2$ and $W^{34}S_2$) at 4 K without a magnetic field, respectively. For 1L $W^NS_2$, the energy band consists of quasi-levels of excitons, trions, and one localized exciton, while for 1L $W^{32}S_2$ and $W^{34}S_2$, two additional quasi-levels form: biexcitons and one localized exciton. In the case of $W^NS_2$, one $M$-phonon is required, while $W^{32}S_2$ and $W^{34}S_2$ require two phonons ($M$ and $\Gamma$) for upconversion. (c–e) A comprehensive graphical demonstration of the aforementioned quasi-levels, accompanied by $K^+$ and $K^-$ valleys under resonant $\sigma^+$ excitation for the 1Ls $W^NS_2$, $W^{32}S_2$, and $W^{34}S_2$, respectively,

depicting the intraband (within $K^+$) and interband (between $K^+$ and $K^-$) scattering. The quasiparticle levels for the exciton, biexciton, triplet and singlet trions, and localized exciton on the $K^+$ valley are represented as $X_{0,\sigma^+}$, $XX_{0,\sigma^+}$, $X^-_{T,\sigma^+}$, $X^-_{S,\sigma^+}$, and $LX_{i,\sigma^+}$, respectively. The radiative emission to the ground state is indicated by a solid arrow, while possible allowed intraband and interband scattering is represented by light brown and black dotted arrows, respectively. The forbidden interband scattering is marked with a red cross.

*Upconversion phenomena*

Upconversion phenomena have been previously studied for 1L $W^NS_2$ under resonance excitation at RT (room temperature), as well as cryogenic temperatures [5,12,13]. However, the occurrence of intraband and interband scattering upon the application of a magnetic field under resonance-polarized PL is not yet fully understood. However, Ma et al. [32] described both intra-valley and inter-valley scattering phenomena by solving the rate equations that describe the valley dynamics for non-resonant magneto-optical properties of excitons and trions in a 1L $WSe_2$. We exploit these two phenomena to discuss the variation of the quasiparticle DOP under magnetic fields.

The upconversion phenomenon occurred differently in the mixed isotopically modified 1L ($W^NS_2$) and pure S isotope-modified 1Ls ($W^{32}S_2$ and $W^{34}S_2$), as shown in Figure 6(a) and (b), respectively. The identification of intermediate energy bands/levels that trigger the upconversion process is essential in order to understand the underlying mechanism [12,13]. The position of the upconversion intermediate state can be obtained through an optical gain analysis. For $W^NS_2$, the estimated gain is ~30 meV (close to the triplet trion peak), while in the case of $W^{32}S_2$ and $W^{34}S_2$, the estimated gain is ~80 meV (close to the first localized exciton peak); it is nearly constant with an applied magnetic field (see Figure 3(d)). Further, Figure 6(a) elaborates on the upconversion process of 1L $W^NS_2$; upon the excitation of resonant light, the valence band (VB) electron at the $K^+$ valley transits to an intermediate virtual state, closely resembling the real trion state ($X^-_T$). From this trion state, the mediating electron (from the VB) absorbs the *M*-phonon ($q_M$ ~ 213 cm$^{-1}$ or 27 meV, refer to Figure S1(a) in the SI) and scatters to the conduction band (CB) in the same valley [68]. The CB electron exhibits strong coupling with the VB hole, eventually leading to the emission of an exciton (~1.989 eV) with an energy greater than the excitation energy. Figure 6(b) exhibits the upconversion process for the pure isotope 1Ls $W^{32}S_2$ and $W^{34}S_2$. Despite the small variation in composition between $^N$S and $^{32}$S, the 1Ls of $W^NS_2$ and $W^{32}S_2$ demonstrate unique upconversion behaviors due to distinct phonon dispersion. In the case of pure isotope 1L samples, the upconversion phenomena are found to be stimulated by two-phonon absorption. The electron in the VB (at the $K^+$ valley) transits to an intermediate virtual energy level, positioned close to the first $LX_1$. At this point, the mediating electron first absorbs the $\Gamma$-phonon *(for $W^{32}S_2$, $q_\Gamma \sim 418$ cm$^{-1}$ or 52 meV, and for $W^{34}S_2$, $q_\Gamma \sim 406$ cm$^{-1}$ or 50 meV; see Figure 2(b) and Figure S1, respectively)*, scatters to the $X_T^-$ band (at the $K^+$ valley), and then further absorbs the M-phonon ($q_M \sim 213$ cm$^{-1}$ or 27 meV *(for both types of pure isotope-modified WS$_2$)* [68]*; see Figure 2(b) and Figure S1*), transiting the mediated electron to the CB. Thereafter, the CB electron couples with the VB hole and emits an exciton with an energy greater than the excitation energy, similar to $W^NS_2$.

*Intra/inter-valley scattering phenomena*

As mentioned in the previous section, the application of magnetic fields induces quasiparticle intraband/interband scattering and valley splitting. The underlying scattering phenomena for 1L $W^NS_2$ are illustrated in Figure 6(c). The analysis of intensity ratios (see Figure 5 and Table S3) reveals that upon applying a magnetic field, excitons at the $K^+$ valley ($X_{0,\sigma^+}$) predominantly scatter to the $K^-$ valley ($X_{0,\sigma^-}$), i.e., interband scattering, which is dominant in $W^NS_2$. Additionally, these excitons also scatter to $X_{T,\sigma^+}^-$, $X_{S,\sigma^+}^-$, and $LX_{1,\sigma^+}$, while a proportion of $X_{T,\sigma^+}^-$ scatter to $X_{S,\sigma^+}^-$ within the same $K^+$ valley, i.e., intraband scattering. Similarly, intraband scattering has been observed from the trion band ($X_{T,\sigma^+}^-$ and $X_{S,\sigma^+}^-$) to $LX_{1,\sigma^+}$, although the scattering probabilities are comparatively smaller, as substantiated by smaller values of the slopes of the intensity ratio (see Table S3 in the SI). Figure 6(d) illustrates the intraband scattering phenomenon of 1L $W^{32}S_2$, where $X_{0,\sigma^+}$ scatters to $X_{T,\sigma^+}^-$, $LX_{1,\sigma^+}$, and $LX_{2,\sigma^+}$. Further, $X_{T,\sigma^+}^-$ shows no trace of scattering to other quasiparticles; however, traces of scattering are observed from $X_{S,\sigma^+}^-$ to $X_{T,\sigma^+}^-$, $LX_{1,\sigma^+}$, $LX_{2,\sigma^+}$ (intraband), and $X_{S,\sigma^-}^-$ (interband). The majority of quasiparticles settle down to the localized states. The scattering process for $W^{34}S_2$ is illustrated in Figure 6(e), and we note that $X_{0,\sigma^+}$ scatters solely to $X_{T,\sigma^+}^-$, $LX_{1,\sigma^+}$ (intraband), and $X_{0,\sigma^-}$ (interband), while $XX_0$ traces no scattering among its bands. It is noteworthy that interband scattering occurs from $X_{T,\sigma^+}^-$ to $X_{T,\sigma^-}^-$, whereas $X_{S,\sigma^+}^-$ exhibits both intraband ($X_{S,\sigma^+}^-$ to $X_{T,\sigma^+}^-$, $LX_{1,\sigma^+}$, and $LX_{2,\sigma^+}$ at the $K^+$ valley) and interband ($X_{S,\sigma^+}^-$ to $X_{T,\sigma^-}^-$ and $X_{S,\sigma^-}^-$ at the $K^-$ valley) scattering. Interestingly, 1L $W^{34}S_2$ shows $LX_{1,\sigma^+}$ scattering to $LX_{2,\sigma^+}$ (intraband) and $LX_{2,\sigma^-}$ (interband). It should be noted that even though the upconversion phenomena in the 1Ls $W^{32}S_2$ and $W^{34}S_2$ occur due to two-phonon absorption, the intraband and interband scattering are quite distinct. Next, we discuss the correlation of the DOP with the intraband and interband scattering through the approach of phenomenological models described in this section.

"$X_0$ DOP" for 1L $W^NS_2$ decreases as the magnetic field increases (see Figure 4(a)) due to the dominant interband scattering from $X_{0,\sigma+}$ to $X_{0,\sigma-}$ and fewer intraband scatterings to $X^-_{T,\sigma+}$, $X^-_{S,\sigma+}$, and $LX_{1,\sigma+}$, as shown in Figure 6(c). This can be understood from the definition of the DOP for a particular quasiparticle. If there is comparatively larger scattering of quasiparticles at the $K^+$ point to other quasi-bands (intra/inter-valley), this will lead to a decreased DOP. Conversely, if the scattering is less predominant, it will result in a higher DOP. Similarly, a minimal decrease in the DOP with the magnetic field for the pure S isotope 1Ls $W^{32}S_2$ and $W^{34}S_2$ is observed. For 1L $W^{32}S_2$, this change is small due to the solely intraband scattering of $X_{0,\sigma+}$ to $X^-_{T,\sigma+}$ and $LX_{i\,(i=1,2),\sigma+}$. However, for $W^{34}S_2$, both intraband scattering and interband scattering, i.e., from $X_{0,\sigma+}$ to $X^-_{T,\sigma+}$, $LX_{2,\sigma+}$, and $X_{0,\sigma-}$, contribute to a relatively larger decrease in "$X_0$ DOP" with the magnetic field. The probability of intra/inter-valley quasiparticle scattering is estimated from the analysis of the slope of the intensity ratio (see Table S3). As mentioned previously, $W^NS_2$ manifests a distinguished phonon spectra due to isotopic disorder. This disorder can enhance intraband and inter-valley scattering processes and reduce the stability of exciton polarization, leading to a decrease in the DOP with an increasing magnetic field [69]. Pure isotope 1Ls result in more uniform phonon interactions [36]. This uniformity can stabilize the exciton states, resulting in minimal change in the DOP with an increasing magnetic field.

"$X^-_T$ DOP" for $W^NS_2$ exhibits a significant increase as the magnetic field increases from 0 to 5 T and thereafter saturates at higher field values (see Figure 4(b)). The increase in "$X^-_T$ DOP" can be attributed to the intraband scattering of $X_{0,\sigma+}$ to $X^-_{T,\sigma+}$, as the amount of triplet trions is enhanced. The initial rapid increase in "$X^-_T$ DOP" for trions in $W^NS_2$ up to 5 T can be attributed to the significant influence of the intraband (or Landau level) splitting, which increases the trion binding energy and enhances polarization [64,70] (see Figure S8(a)). However, beyond 5 T, the binding energy of trions reaches a stable state, where additional magnetic field strength does not significantly affect it. This stabilization leads to a gradual saturation in "$X^-_T$ DOP," as the trions are already highly polarized, and further increases in the magnetic field contribute minimally to additional polarization.

Similarly, the increase in "$X_T^-$ DOP" for W$^{32}$S$_2$, with the increase in the magnetic field, can be attributed to the intraband scattering of $X_{0,\sigma^+}$ and $X_{S,\sigma^+}^-$ to $X_{T,\sigma^+}^-$. In contrast, "$X_T^-$ DOP" for W$^{34}$S$_2$ exhibits a decrease, reaching ~-27% with an increase in the magnetic field, due to the interband scattering of $X_{T,\sigma^+}^-$ and $X_{S,\sigma^+}^-$ to $X_{T,\sigma^-}^-$. The increase in "$X_T^-$ DOP" in W$^{32}$S$_2$ with an increasing magnetic field can be due to stronger electron-phonon interactions, weaker hyperfine interactions, and efficient scattering mechanisms that help align trion spins. Conversely, the decrease in "$X_T^-$ DOP" in W$^{34}$S$_2$ can be attributed to weaker electron-phonon interactions, stronger hyperfine interactions, and more complex scattering mechanisms due to heavier isotopes that disrupt trion spin alignment [29,58,60].

Furthermore, "$X_S^-$ DOP" for W$^N$S$_2$ initially increases with an applied field from 1 to 5 T (Figure 4(c)), which can be attributed to the intraband scattering of $X_{0,\sigma^+}$ and $X_{T,\sigma^+}^-$ to $X_{S,\sigma^+}^-$, but the scattering slows down at higher fields, resulting in a nearly saturated DOP. This follows the same behavior and reasoning as the case of "$X_T^-$ DOP" for W$^N$S$_2$. In W$^{32}$S$_2$, "$X_S^-$ DOP" decreases with increasing magnetic fields. This is because $X_{S,\sigma^+}^-$ exhibits intraband scattering to $X_{T,\sigma^+}^-$ and $LX_{i\,(i=1,2),\,\sigma^+}$, as well as interband scattering to $X_{S,\sigma^-}^-$ (see Figure 6(d)). Similarly, in W$^{34}$S$_2$, "$X_S^-$ DOP" decreases with increasing magnetic fields due to intraband scattering from $X_{S,\sigma^+}^-$ to $LX_{i\,(i=1,2),\,\sigma^+}$ and interband scattering to $X_{S,\sigma^-}^-$ and $X_{T,\sigma^-}^-$. This behavior can be understood through the interband and intraband (Zeeman) splitting strength (slope or *g*-factor), which indicates that singlet trions exhibit strong coupling with the magnetic field (see Figure S8 and Tables S4 and S5). This strong coupling leads to the increased scattering of singlet trions, resulting in a decrease in their "$X_S^-$ DOP" when a magnetic field is applied [71]. Previously, similar trends of exciton and trion DOPs under fields have been reported for the 1Ls W$^N$S$_2$ and WSe$_2$ [33,34,57].

Interestingly, "$XX_0$ DOP" is nearly constant under varying magnetic fields (see Figure S7(a)) because $XX_0$ exhibits neither intraband nor interband scattering for pure isotope-modified WS$_2$ 1Ls. Additionally, in all WS$_2$ 1Ls, a high probability of the intraband scattering of quasiparticles (exciton and trions) to the *LX* state is observed with an increase in the magnetic field, resulting in the increase of "*LX* DOPs." At low temperatures, LX states in WS$_2$ are considered stable due to the strong Coulomb interactions, localized potential wells in the crystal lattice, lower energy configurations, longer lifetimes, and experimental evidence confirming their existence [48,72]. Further, at higher magnetic fields, increased scattering probabilities can affect the contribution and interaction of quasiparticles to stabilize *LX* states by promoting relaxation into these energetically favorable states.

The differences observed in $WS_2$ properties between $^NS$-, $^{32}S$-, and $^{34}S$-derived samples stem from the combined effects of phonon spectrum variations and isotopic disorder [39,59], highlighting the intricate relationship between these factors in determining the material's characteristics.

## 4. Conclusions

In summary, we investigated the influence of varying S isotopes in 1L $WS_2$ on PL emission under resonance excitation, as a function of the temperature, magnetic field, and light polarization. By analyzing the DOP and intensity ratios of PL emission corresponding to different excitonic transitions, we observed distinct variations among $WS_2$ 1Ls composed of different S isotopes. Our analysis of intensity ratios and intra/interband (Zeeman) splitting led to a proposed phenomenological model involving intraband/interband scattering and upconversion processes. Notably, upconversion in 1L $WS_2$ with natural S isotopic abundance involves single-phonon absorption, whereas 1L $WS_2$ composed of pure $^{32}S$ and $^{34}S$ isotopes exhibits two-phonon absorption processes. Moreover, the different S isotopes exhibit unique scattering behaviors: $W^{32}S_2$ shows strong intraband scattering, $W^{34}S_2$ exhibits robust interband scattering, and $W^NS_2$ displays contributions from both. Interband scattering notably results in a more negative DOP, while intraband scattering has the opposite effect. Through intraband scattering processes, quasiparticles are redistributed into localized electronic states, with their degree of localization influenced by the isotope composition. This effect is consistently observed across all isotopic samples, highlighting its fundamental nature. Additionally, the upconversion emission, tunable via isotope variation, enhances the overall light emission efficiency. These findings underscore the potential of S isotope engineering to modulate the properties of 1L TMDs, presenting new opportunities for advancing optoelectronic and photonic devices based on 1L $WS_2$.

## Acknowledgements:

The work was supported by the Czech Science Foundation, project no. 24-11772S and Ministry of Education, Youth and Sports of the Czech Republic, project QM4ST, no. CZ.02.01.01/00/22_008/0004572.

We also acknowledge the support in the AFM and low-temperature environments provided by the Research Infrastructures NanoEnviCz and MGML, supported by the Ministry of Education, Youth and Sports of the Czech Republic under projects no. LM2015073 and LM2023065, respectively.